# Alphanetv4: Alpha Mining Model

**Wenjun Wu**
University of Illinois Urbana-Champaign

**Abstract**
As AI and deep learning have become hot spots in the 21st $^{century}$, they are widely used in the current quant market. In 2020, Huatai Securities constructed deep-learning-based AlphaNet for stock feature extraction and price prediction. At present, it has developed to the 3rd $^{version}$ and has formed a great influence in the market.

However, the AlphaNet has some problems, such as underfitting caused by short sequence length of feature extraction, insufficient diversity of feature extraction, high complexity, instability of random sampling, which lead to the poor performance. So this paper proposes AlphaNetV4 to solve them. The main contributions of this paper are: 1) Increased the length of the sequence and reduced the step size of the extraction layer to improve the fitting effect; 2) Reduced the relevance of original input; 3) Used Spearman correlation coefficient to design dropout layer instead of random sampling to enhance the stability of feature extraction; 4) Applied Bi-LSTM to enrich the extraction layer, and Transformer to enhance the learning ability of the model. In addition, this paper also uses CNE5 Barra to redesign the fitting target, and optimizes the training process by modifying the training weight and using sharp EarlyStopping. This paper compares the performance between AlphaNetV4 and the previous AlphaNets. It verifies that increasing the sequence length can reduce the loss from 0.5 to 0.3, reducing the correlation of input can reduce the loss to 0.25, using Spearman Dropout can cut the computational complexity without damaging the accuracy, and that Transformer can reduce the loss to less than 0.1. Further, this paper conducts the back test to show that AlphaNetV4 has increased the annual excess return by about 7% - 10%. Finally, this paper provides suggestions on the future development of quant trading.

# 1. introduction
## 1.1 Research background and significance
### 1.1.1 Background

Stock market investment is one of the hot topics in the financial field. With the sharp increase in the number of stockholders and the amount of capital, the stock market has been in a state of ups and downs and unpredictability. Researchers have difficulty in giving a satisfactory explanation for such phenomena using traditional financial theories. Quantitative finance is to use statistical and quantitative methods to establish strategies and obtain the invariance contained in variable numbers through the powerful computing power of computers .

Generally speaking, the returns of quantitative investment come from stock selection and timing. Among them, timing is to produce technical aspects through the exploration of fundamentals, that is, the factors of the stock market. Factors are very important components in quantitative stock portfolio management. Only good factors and combining them in the right way can build a high-quality model, so factor analysis is extremely important. In our daily life, we will find that the concept of "academic tyrants" does not only exist in a single subject. Students with good grades in a certain subject will also perform very well in other subjects. This phenomenon was first discovered by the British psychologist Spearman. Later, he speculated whether there are some common factors that affect the occurrence of this phenomenon, and thus proposed the concept of factor analysis. Nowadays, factor analysis has also become the main idea of dimensionality reduction and finding representative factors in complex variables.

The main point of stock selection is investment portfolio, the purpose is to diversify risks. Investment portfolio in the stock market is to consider how to combine risky assets. Since the risk-return of any two assets with poor or negative correlation will be greater than the risk-return of individual assets, continuously combining assets with poor correlation can keep the effective frontier of the portfolio away from risks.

Considering the importance of factor analysis and portfolio in stock market and quantitative finance, this paper studies the problems of factor mining, stock price prediction and portfolio based on deep learning method.

### 1.1.2 Research significance

As big data and artificial intelligence have become hot topics in the 21st century, machine learning has become widely known and has been widely used in various fields. As machine learning becomes more mature, researchers have also applied it to stock price prediction in the field of financial investment . The reason why machine learning can perform better in stock price prediction is that the increasingly complex stock market has caused the linear model to reach a bottleneck in development, while the nonlinear model provides more possibilities for quantitative trading. Deep learning has a broad application prospect in quantitative trading due to its good fitting ability for nonlinear relationships and has gradually become the main exploration direction of quantitative trading. Deep learning mainly uses factor mining, graph neural networks, social media and other unstructured data, knowledge graphs and time series models to improve these problems in quantitative trading.

By stacking a large number of parameter matrices, neural networks can extract valuable information for conclusions from data that is not very readable for humans. In addition, due to the strong plasticity of neural networks, they can be improved and perfected in many directions, thereby continuously optimizing the prediction accuracy . Therefore, studying the development of neural networks in quantitative factor mining and stock selection is of great significance to the development of quantitative finance and even the entire artificial intelligence.

Although many financial companies have made great achievements in this field in recent years, the highly confidential nature of the financial industry makes it difficult for the general public to learn about the quantitative strategies developed by funds and investment institutions . Therefore , this article will provide a series of processing methods and improvement directions for deep learning in the field of quantitative investment.

### 1.2 Research innovation

Neural networks ( deep learning ) mainly include convolutional neural networks (CNN) , recurrent neural networks (RNN) , attention mechanisms, etc. Due to the structural characteristics of CNN and RNN, they are networks built by combining knowledge in specific fields, and they may not work well if they are separated from the original fields. Therefore, when they are applied to the stock market for quantitative multi-factor stock selection and mining, various problems will arise. For example, the local perception operation of the convolution kernel of CNN is related to the arrangement of data, but stock factor data does not have a fixed arrangement, which makes the local perception feature meaningless in pure numerical data.

Based on the characteristics of these neural network structures and the characteristics of the stock market, Huatai Securities proposed the current SOTA of deep learning quantitative investment - AlphaNetV 3 , which uses a combination of custom feature extraction layers and deep neural networks to perform factor mining and price prediction. However, this paper found through empirical research that AlphaNetV 3 has the following problems:

1. The feature extraction sequence length is too short, which leads to the problem of underfitting of the model.
2. The correlation between the original inputs is too high, which reduces the diversity of feature extraction and increases invalid computational overhead.
3. The randomness of random sampling increases the instability of the model during the final fitting.
4. The neural network is too single, resulting in poor fitting effect.

To address the above problems, this paper proposes an improved algorithm AlphaNetV4, which solves the above problems through the following methods:

1. This paper increases the sequence length and reduces the extraction layer step size to improve the fitting effect.
2. This paper reduces the correlation of the original input by taking the growth rate.
3. This paper designs the DropOut layer through the Spearman correlation coefficient to replace random sampling.
4. This paper replaces the original double-layer LSTM and double-layer GRU designs with Bi-LSTM to enrich the design of the extraction layer, and enhances the learning ability of the model through residual optimization of Transformer.

In addition to solving the current defects of AlphaNetV 3 , this paper also enhances the performance of the model in real stock trading through the following methods:

1. is used to perform Barra factor style stripping to change the fitting target, making factor mining more pure.
2. Adding the weight of multi-head samples in the loss function makes the model training go in the direction expected by this article.
3. Using Sharp Early Stopping to reduce the number of model training cycles makes the model more suitable for real stock investment.

## 2. Literature review and theoretical basis of AlphaNet model

This chapter first lists the development history of artificial intelligence in the field of quantitative investment, and reviews several classic deep learning models : CNN , RNN and Attention . Neural networks are widely accepted, recognized and deeply studied in the fields of image recognition and speech recognition because they can achieve end-to-end learning and prediction, have strong scalability and are applicable in various fields . However, neural networks need to combine the understanding of network structure and certain domain knowledge to design the network structure in a targeted manner. Therefore, understanding the principles of neural network structure is the basis for such models to be used in stock price prediction, identify investment opportunities, reduce risks and obtain excess returns .

### a. Domestic and foreign literature review

The application of neural networks in the field of quantitative investment was in 1988 when White [1] used neural networks to predict the stock return rate of IBM, but it was not successful in this regard. It was not until 2003 that scholar De Lisle [2] achieved satisfactory results in using neural networks to predict stock index prices. In the same year, Moody's proposed the RRL model [3] to bring the idea of reinforcement learning into quantitative investment. Later, other scholars integrated genetic algorithms and Bayesian regularization into the quantitative investment of neural networks and the improvement of traditional momentum models.

Domestic research on the development of neural network quantitative investment tends to focus on empirical research. Since 2004, Professor Cui Jianfu's research on the superiority of BP neural network model over GARCH model [4] has opened the door for domestic scholars to study neural network quantitative investment. In 2008, an article in the Journal of Tsinghua University entitled "Complex Financial Time Series Prediction Based on Least Squares Support Vector Machine" [5] improved the problem of high computational overhead in big data analysis and laid the foundation for efficient training of neural networks. In 2012, Professor Yao Xiao also introduced the concept of fuzzy membership [6] on support vector machines, overcoming the interference of white noise and making the model prediction more accurate. In 2014, researchers began to apply random forests to equal-frequency stock data. In 2015, data dimensionality reduction began to receive attention: some studies used neural networks and genetics to reduce dimensionality, while some researchers used PCA to embed it into some existing machine learning algorithms [7]. Since 2016, most domestic quantitative companies have begun to use Adaboost ing technology [3] and produced a large number of research reports.

The use of deep learning for quantitative investment in China is a product of the rapid development of deep learning in recent years. For example, in 2017, researchers applied it to the representation and learning of financial signals to assist decision-making; in 2018, reinforcement learning was applied to forecasting and investment portfolio strategies, etc.

Huatai Securities proposed AlphaNet [15] in 2020, bringing the idea of CNN feature extraction into the quantitative field. This paper improves the original model based on AlphaNet and proposes AlphaNetV 4.

b. Theoretical basis of AlphaNet model

i. CNN

Convolutional Neural Network (CNN) is the most influential model in the field of computer vision research and application, and is also a milestone achievement at the turning point of deep learning methods. It was first proposed by Hinton [8]. In 2012, ImageNet, proposed by Hinton's team with the help of the convolution idea, achieved amazing accuracy in image classification, which opened the CNN craze. The idea of CNN is similar to the visual nerve. For humans, the visual nerve is a tool for the eyes and the brain to exchange information. The visual nerves work together, each responsible for a small area of the visual image and exchanging information. Finally, through information integration, feature abstraction, and transmission of abstract features to the brain, humans can use vision to generate cognitive abilities. Convolutional neural networks imitate this principle. Multiple convolution kernels extract features from local areas of the image, and aggregate them through the pooling layer to obtain features with a higher level of abstraction. Finally, the fully connected layer is used to synthesize features. The figure below shows a classic convolutional neural network model.

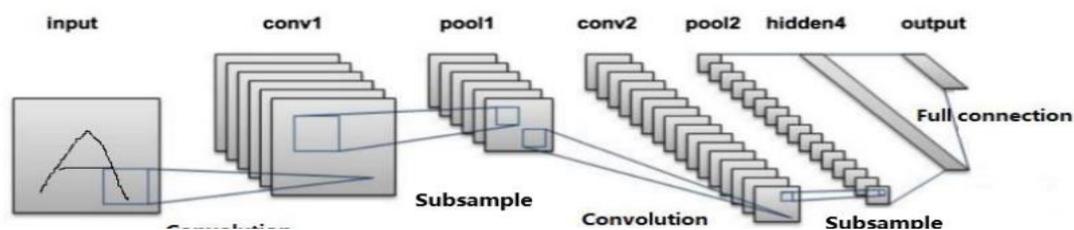

图 2-1 CNN 结构

The most important part of CNN is the convolution kernel. As shown in the figure below, the convolution kernel is a network layer with optimizable weights and biases. Through the action of multiple layers of convolution kernels, the edges, facial features and facial features can be gradually extracted from the image, and finally face recognition can be completed.

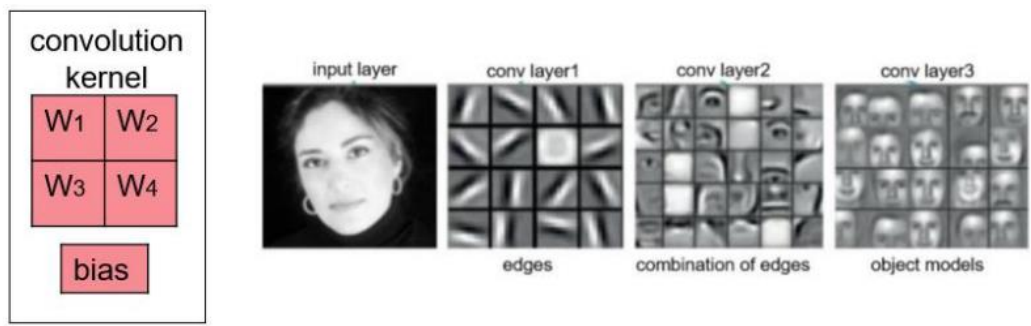

Figure 2-2 CNN recognition mechanism

CNN makes end-to-end image recognition possible, avoiding artificial feature extraction. Based on the feature extraction principle of image recognition, the convolution kernel pooling network structure is cleverly designed, reflecting the powerful feature learning ability of deep neural networks.

However, for CNN, the local perception operation of the convolution kernel is related to the arrangement of data. This fixed arrangement does not exist in pure numerical data, which makes the local perception feature meaningless in pure numerical data. Secondly, the operation of the convolution kernel is actually a weighted average and maximum and minimum operation on the data in the area. This method may make it difficult to extract the features contained in stocks.

ii. **RNN and LSTM**

Neural Network (RNN) is a neural network that takes a sequence as input and has a directed recursion, making all nodes connected in a chain. The time sequence construction of RNN makes it very robust when learning the nonlinear features of the sequence. There are generally four ways for RNN to model the sequence:

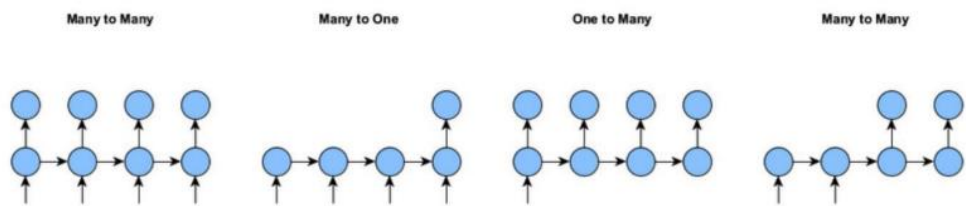

Figure 2-3 RNN modeling method

In theory , RNN can connect previous information with the current task , but if it is a more complex task, there is a situation where the distance between useful information and the information to be processed is far, and RNN, whose recursive method is too simple, is difficult to extract more complex features in stock data, and the derivation task may fail. As shown in the following formula, RNN updates parameters by continuous derivation in a long program sequence. If tj is very large, that is, j is far away from the t time step, and $\sigma'W^h$ > 1, there will be a gradient explosion problem; when $\sigma'W^h$ < 1, there will be a gradient vanishing problem. This type of problem leads to the defect that RNN cannot be "long-term dependent".

$$\frac{\partial E}{\partial W^h} = \sum_{i=1}^{t} \frac{\partial E}{\partial y^t} \frac{\partial y^t}{\partial h^t} \frac{\partial h^t}{\partial h^i} \frac{\partial h^i}{\partial W^h} \quad (\text{in } \frac{\partial h^t}{\partial h^i} = \prod^{t-j} \sigma'W^h \ )$$  Formula 2-1

LSTM is to solve this "long dependency" problem . Long Short Term Memory networks (LSTM )  [9], introduced by Hochreiter and Schmidhuber in 1997 , use forget gates, input gates, and output gates to control the cell state, as shown in the figure below.

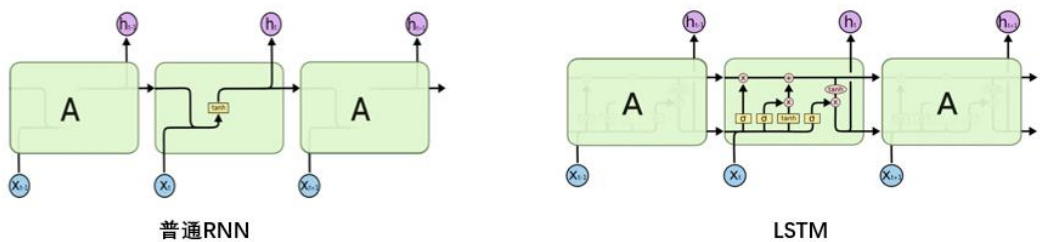

Figure 2-4 Comparison between RNN and LSTM

First, LSTM needs to decide which information to discard. This part of the operation is done through a "forget gate" composed of a sigmoid activation function. It looks at the information of $h_{t-1}$ and $x_t$ and uses sigmoid activation to output the obtained information as a vector between 0 and 1. This vector determines $C_{t-1}$ which information in the cell state needs to be retained and which information needs to be discarded. The forget gate formula and structure are as follows:

$$f_t = \sigma\left(W_f \cdot [h_{t-1}, x_t] \ + \ b_f\right)$$  Formula 2-2

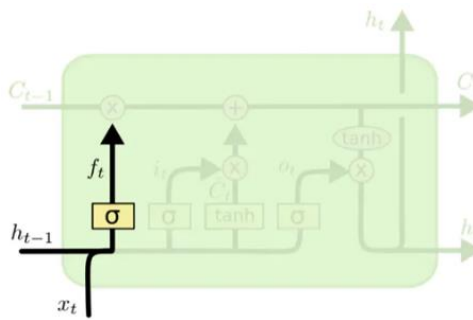

Figure 2-5 Forget gate structure

The next step is to decide what new information to add to the cell state. This step is divided into two steps. First, use $h_{t-1}$ and $x_t$ pass an operation called input gate to decide which information to update. Then use $h_{t-1}$ and $x_t$ pass a tanh layer to get new candidate cell information $\bar{c}_t$, which may be updated to the cell information. The formula and structure are as follows:

$$i_t = \sigma(W_i \cdot [h_{t-1}, x_t] + b_i)$$   Formula 2-3

$$\tilde{C}_t = \tanh(W_C \cdot [h_{t-1}, x_t] + b_C)$$   Formula 2-4

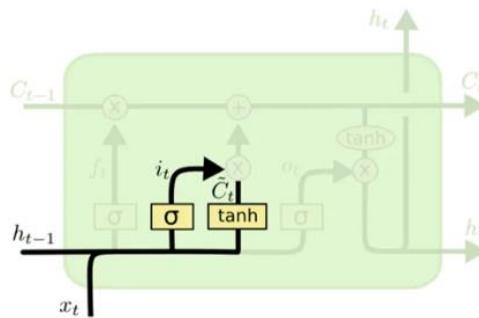

Figure 2 - 6 -input gate structure

Next, we will update the old cell information $\bar{c}_{t-1}$ to the new cell information $\bar{c}_t$. The update rule is to forget part of the old cell information through the forget gate, and add $\bar{c}_t$ part of the candidate cell information through the input gate to get the new cell information $c_t$. The formula and structure of the update operation are as follows:

$$C_t = f_t * C_{t-1} + i_t * \tilde{C}_t$$   Formula 2-5

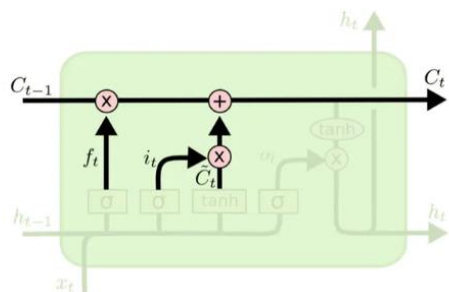

Figure 2-7 Conveyor belt structure

After completing the operations on the "conveyor belt", you need to use $h_{t-1}$ and $x_t$ to determine the features that need to be output. Here, $h_{t-1}$ the and $x_t$ are combined into the same vector, activated by the sigmoid layer to obtain the judgment condition $o_t$, and the combined vector is also activated by the "output gate" composed of the tanh layer, and finally the vector is $o_t$ multiplied by the judgment condition to obtain the output vector. The steps are as follows:

$$o_t = \sigma(W_o [h_{t-1}, x_t] + b_o)$$   Formula 2-6

$$h_t = o_t * \tanh(C_t)$$   Formula 2-7

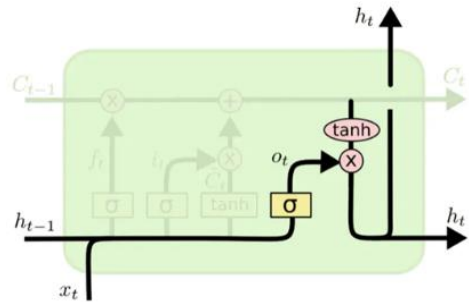

Figure 2-8 Output gate structure

LSTM takes into account the structure of time series to a great extent and effectively weakens the problem of long-range dependency. In addition, LSTM avoids the problem of gradient vanishing and explosion through its own complex structure. This article is based on the Many to One and Many to One way to synthesize factors.

iii. **Attention and Transformer**

The attention mechanism was proposed by the Bengio [10] team in 2014 to solve the long-term forgetting problem in machine translation Seq 2 Seq (an application of RNN). In recent years, it has been widely used in various fields of deep learning.

As shown in the figure below, in the original RNN-based model, the output is directly the last state $h_m$, while Attention has an additional "look back" mechanism. After RNN finishes working, Attention starts working. Where $S_0$ is the last state of RNN $h_m$, and each previous hidden state h must be retained. Here, $S_0$ the correlation with each result must be calculated and the result is recorded as $a_i$, called weight Weight. The way to calculate the correlation is to connect and $h_i$ and $S_0$, use matrix W to calculate the product of the connected vector, after tanh transformation, and finally calculate the inner product with matrix v to get the real number $\overline{\alpha_i}$, where matrix V and W are parameters that need to be learned. After the initial calculation $\overline{\alpha_i}$, use Softmax transformation to get the final $a_i$.

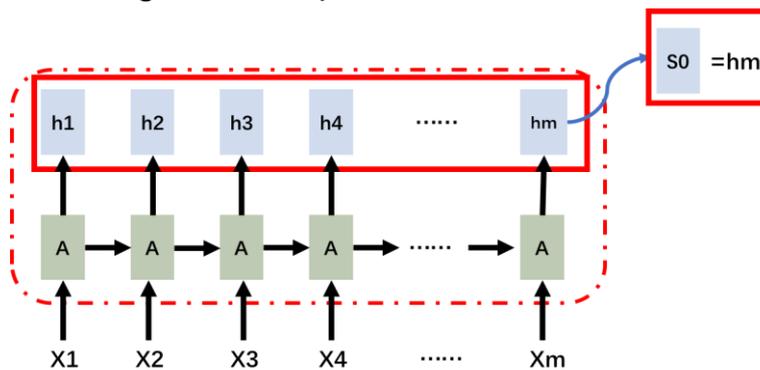

Figure 2-9 Attention mechanism

Use these weights $a_i$ to perform a weighted average on the m weights, and the result is a vector $c_0$ called context vector, each of c which corresponds to a state s, and these c are the final outputs.

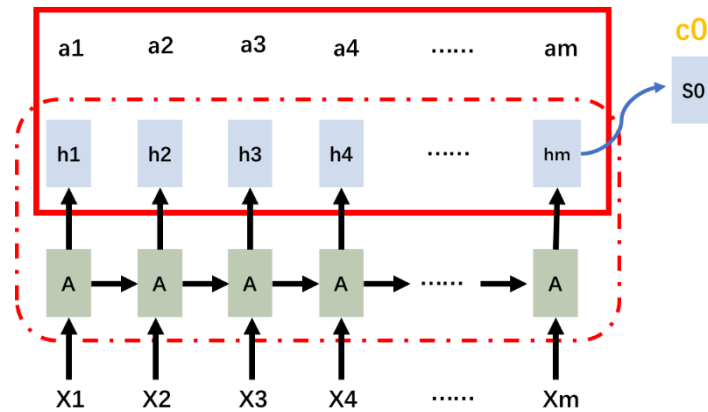

Figure 2-10 Attention update method

At NIPS 2017, a paper titled "Attention is All You Need" [20] proposed the current SOTA model based on Attention - Transformer. It abandons the traditional CNN and RNN, and the entire network structure is composed entirely of Self-Attention and Feed Forward Neural Network mechanisms. Due to the complex structure of Transformer, only the Transformer used in this paper is introduced here. Encoder structure.

The Encoder is a Multi-Head built by multiple Self-Attention The structure of Attention. One of the Self-Attention structures and calculation methods is as follows:

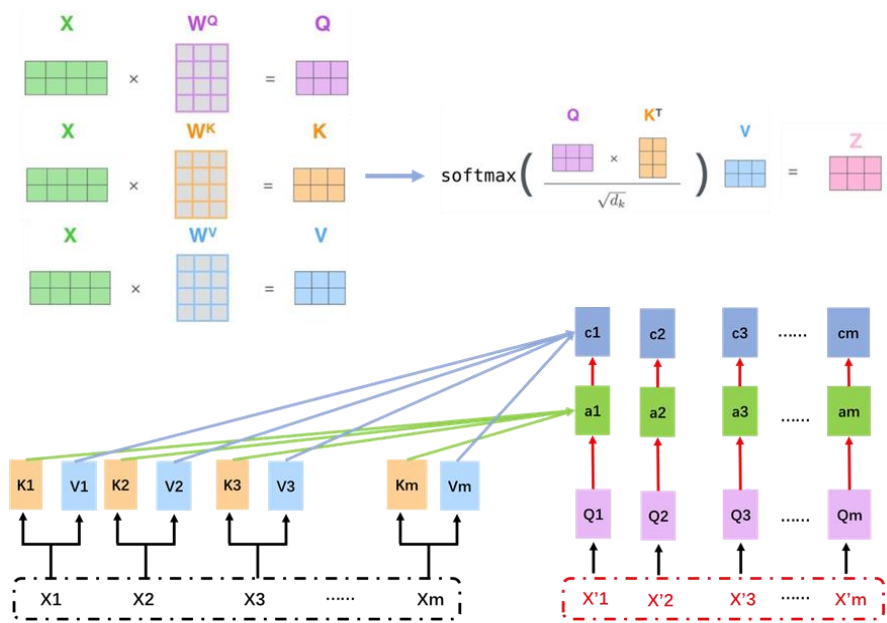

Figure 2-11 Transformer structure and update method

As mentioned above, the two-dimensional information matrix is transformed by $W^q$ three matrices, $W^v$ Q $W^k$, K, and V, respectively $W^q$, to obtain the required three tensors. In self-attention, the length of the input and output vectors is the same, both are m-dimensional, so after mapping by $W^v$, $W^k$ we will get m Q, K, and V vectors. For each input X, calculate the score (where score = $Q * K^T$), in order to stabilize the gradient, the Encoder uses the normalization of the score, that is, dividing by $\sqrt{d_k}$, performing a softmax transformation on the score to make it a probability distribution, and multiplying this probability distribution by the Value value to obtain the weighted score Z of each input vector.

After building a Self-Attention, stacking multiple Self-Attention layers together becomes a Multi-Head Attention, each Attention does not share parameters. A Self-Attention has 3 parameter matrices, so the Multi-Head layer accumulated by m Self-Attention layers Attention has 3m parameter matrices. In order to ensure that the input and output tensors are of the same size, the Transformer Encoder is in Multi-Head Based on Attention, the skip-connection structure in the residual network is adopted (corresponding to Add in the figure below). & Normalize), the purpose is to solve the degradation problem in deep learning, and the final Encoder performance is shown in the following figure:

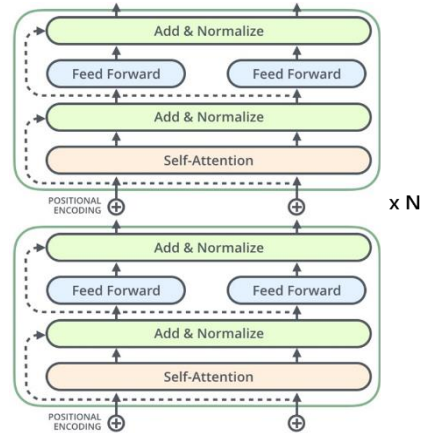

Figure 2-12 Multi-Head Attention

### iv. AlphaNet

AlphaNet [18] is to reconstruct and redefine the commonly used neural network so that the neural network model can be applied in the real stock market. This paper improves the prediction accuracy and stability by reconstructing and stacking complex models, and avoids overfitting.

Among them, AlphaNet is designed to effectively extract features from the original volume and price data. It organizes the data into a two-dimensional "data picture" by imitating the CNN method, and extracts the features required by this article through its own defined feature extraction layer. AlphaNet defines the following feature extraction layers: Pearson correlation coefficient layer, Z-score layer, linear attenuation layer, growth layer, and standard deviation layer. After passing through the feature extraction layer, this article will standardize the data in the form of daily data sections. The purpose of standardization is as follows:

First, standardization can speed up the convergence of the model and effectively avoid gradient disappearance.

Second, standardization can improve the generalization ability of the model. The scaling factor can effectively avoid excessive or small weights caused by different dimensions. It can efficiently identify neurons that contribute less to the network and weaken or eliminate these neurons through activation functions.

Third, standardization can improve the robustness of training and solve the problem of different parameter dimensions under different distributions.

In the time series forecasting model, traditional ARIMA is a process of converting a time series into a stationary state through a residual sequence. However, the premise of the residual sequence model is that the covariance between the model mean and its observations is fixed. However, in the real stock market, this premise does not exist, and even if it does, it is not suitable for long-term time series forecasting. Today, due to the development of machine learning and neural networks, they can identify more complex nonlinear patterns in time series through the stacking of a large number of neurons and nonlinear functions . Therefore, in AlphaNetV3 released in 2021 [17], Lin Xiaoming's team adopted LSTM, which is more outstanding in time series forecasting. Its gated structure can better capture the dependencies between long-range data.

# 3. Transformer-based AlphaNetV 4 model construction

## a. AlphaNetV 4 construction ideas

Huatai Financial released a report titled "Quantitative Factor Mining Based on Genetic Programming" on June 10, 2020 [15], pointing out that genetic programming is an excellent feature generation tool that can be used to generate massive mining factors. As a result, they created an automatic factor mining algorithm - AlphaNet. This algorithm uses multiple operator functions as custom network layers for feature extraction. This paper improves the original version based on AlphaNet and invents AlphaNetV 4. The figure below shows the global structure of AlphaNetV 4. Next, this paper analyzes each element in it.

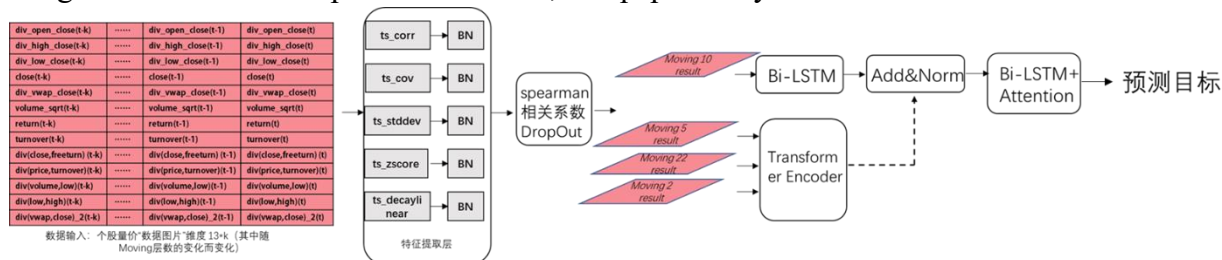

Figure 3-1 AlphaNetV4 structure

## b. Model input and output design

### i. AlphaNetV4 's independent variables

AlphaNet follows the CNN approach and organizes the volume and price data of individual stocks into "data images" and inputs them into the network. The following table explains the meaning of each input variable. The figure below shows a "data image" composed of the volume and price data of a stock from t-4 to t, with the stock 's return rate R (t+10) for the next 10 days as the label. In a single day's trading data, there are about 4,300 stocks, so 4,300 "data images" and their corresponding labels can be generated every day. It is worth noting that the volume and price data images of individual stocks do not have to be arranged in the order of the figure below, and can be arranged in any way.

Table 3-1 AlphaNetV 4 factor description

| name | definition |
| --- | --- |
| div_open_close | open) – closing price ( close) of a certain day divided by the closing price ( close) . |
| div_high_close | The highest price (high ) – closing price ( close) of a certain day divided by the closing price ( close) . |
| div_low_close | The lowest price (low ) – closing price ( close) of a certain day divided by the closing price ( close) . |
| close | Closing price on a particular day. |
| div_vwap_close | Volume Average Price (VWAP) for a particular day – Closing Price ( Close) divided by Closing Price ( Close) . |

Table 3-1 AlphaNetV4 factor description

| | |
| --- | --- |
| volume_sqrt | Take the eighth root of the stock's trading volume on a particular day . |
| return | The growth rate of a stock's closing price on a certain day relative to the previous day. |
| turnover | The turnover rate of a stock on a particular day. |
| div_close_freeturn | The closing price on a certain day (close) divided by the free turnover rate of the stock on that day (free turnover). |
| div_price_turnover | The opening price of a stock on a certain day (open) * free turnover rate (free turnover)/( turnover ) * closing price (close). |
| div_volume_low | The trading volume on a certain day (volume) divided by the lowest price (low) of the stock on that day. |
| div_low_high | The lowest price (low) on a particular day divided by the highest price (high) of the stock on that day. |
| div_vwap_close_2 | The volume average price (VWAP) of a stock on a certain day divided by the closing price (close) of the stock on that day. |

**ii. AlphaNetV4's dependent variable: 10- day return after stripping out the Barra style factor**

In order to mine factors with incremental information, this paper adds a factor neutralization mechanism to the objective function of AlphaNet. The neutralization mechanism adopted in this paper is the Barra model [13], which uses the style factor stripping method to make the portfolio have no exposure to factors other than the target factor, so that the new objective function finally fitted can guide the model to mine factors with low correlation with the Barra style factor and reduce style exposure. Barra factor analysis has the following three purposes:

First, factor analysis can obtain the correlation coefficient between individual stock returns. Since the number of individual stocks in the market is huge, if the correlation coefficient is calculated using the return sequence of the individual stocks themselves, the return matrix will be rank-deficient and irreversible. Barra factor analysis solves this problem by decomposing the returns of individual stocks into some common factors.

Second, the Barra model can achieve a dimensionality reduction effect. The number of stocks in the stock market is far greater than the number of factors. Decomposing the rate of return into factors with a relatively small number of stocks can greatly reduce the workload and improve the accuracy of performance attribution and prediction.

Third, the Barra model can perform risk attribution analysis for asset portfolios. In the quantitative field, factors are divided into Alpha factors and risk factors. The former has a stable predictive ability for stock returns, while the latter is a factor that quantitative researchers want to strip out and control. The Barra model extracts risk factors to determine which factors can explain the fluctuations in the return of an investment portfolio.

The MSCI Barra China Equity Model used in this article (CNE5) [13]. This model decomposes the return of the asset portfolio into country factors, multiple industry factors, and multiple style factors. Assume that there are N stocks, P industries, and Q style factors in the market. At any given time point, construct a regression:

$$\begin{bmatrix} r_1 - r_f \\ r_2 - r_f \\ \vdots \\ r_N - r_f \end{bmatrix} = \begin{bmatrix} 1 \\ 1 \\ \vdots \\ 1 \end{bmatrix} f_C + \begin{bmatrix} X_1^{I_1} \\ X_2^{I_1} \\ \vdots \\ X_N^{I_1} \end{bmatrix} f_{I_1} + \cdots + \begin{bmatrix} X_1^{I_P} \\ X_2^{I_P} \\ \vdots \\ X_N^{I_P} \end{bmatrix} f_{I_P} + \begin{bmatrix} X_1^{S_1} \\ X_2^{S_1} \\ \vdots \\ X_N^{S_1} \end{bmatrix} f_{S_1} + \cdots + \begin{bmatrix} X_1^{S_Q} \\ X_2^{S_Q} \\ \vdots \\ X_N^{S_Q} \end{bmatrix} f_{S_Q} + \begin{bmatrix} u_1 \\ u_2 \\ \vdots \\ u_N \end{bmatrix}$$

Figure 3-2 Barra equation

where $r_n$ is the rate of return of the nth stock, $r_f$ is the risk-free rate of return. $X_n^{I_P}$ is the exposure of stock n to the industry $I_P$, $X_n^{I_P} = 0$ means that the stock does not belong to this industry, $X_n^{I_P} = 1$ means that the stock belongs to this industry. $X_n^{S_q}$ is the exposure of stock n to the style factor $S_q$, and its value has been standardized. $u_n$ is the specific return of stock n, which is the part of the excess return that cannot be explained by the factor. $f_c$ is the factor return of the country factor; is $f_{I_p}$ the factor return of $f_{S_q}$ the industry factor; $I_P$ is $S_q$ the factor return of the style factor. There are 1 country factor, 10 style factors and 32 industry factors in CNE5. The following will give a detailed introduction to these three factors:

Among them, the country factor is the first attempt to be added by CNE5. It represents the market portfolio weighted by the circulating market value. In A-shares, the factor exposure of each stock to the country factor is 1.

Style factors exist for two purposes: (1) to observe market style. (2) to observe the source of portfolio risk and return. When regressing, the factors must be standardized so that the standard deviation of each style factor is 1. $S_n$ Represents the weight of the circulating market value of stock n. There is an assumption that the market is neutral to all style factors, and rational investors will ensure that the sensitivity of the style factors of their own investment portfolios is the same as that of the entire market. For specific CNE5 model style factors, please refer to Appendix 1 at the end of the article. Therefore, the constructed investment portfolio satisfies the normal distribution with a mean of 0 and a standard deviation of 1 on all style factors, that is, the following formula exists:

$$\sum_{n=1}^{N} s_n X_n^{S_q} = 0, \quad q = 1, \ldots, Q \qquad \text{Formula 3-1}$$

The industry factor describes whether a stock belongs to this industry. Generally speaking, the 32 industry factors are mutually exclusive. That is, assuming that stock I belongs to the materials industry, its industry factor exposure in the materials industry is 1, and its factor exposure in other industries is 0.

In the T-period regression, CNE5 uses the factor exposure values at the beginning of the period and the stock's return rate in period T for regression. This paper uses daily returns, so the regression should also be daily.

In the test , after adding the neutralization mechanism to the objective function of AlphaNetV 4 , the excess return of the model decreased, but the stability was greatly improved, and the control of drawdown and volatility was significantly improved . It is worth noting that the neutralization mechanism of the loss function is not only applicable to AlphaNet, but also to any neural network stock selection model. It is a universal method . The specific neutralization method is as follows:

Assume that the actual rate of return is y, and the rate of return predicted by the model is $y(x)$, the conventional minimum mean square error loss function (MSE) can be expressed as follows:

$$\text{Loss} = \text{MSE}(y, y(x)) \qquad \text{Formula 3-2}$$

Since the neutralization mechanism is not involved, the model trained by the above formula is usually highly correlated with the existing factors. This paper improves the objective function. Before calculating the MSE, the Barra style factor is first regressed to obtain the residual $y(x)y(x)_{res}$, and then the residual is used as the loss function with $y$ the calculated MSE, as shown in the following formula:

$$\text{Loss} = \text{MSE}(y, y(x)_{res}) \qquad \text{Formula 3-3}$$

The loss function with the neutralization mechanism can guide AlphaNet to explore factors with low correlation with Barra style factors, reduce style exposure, and make the prediction results of AlphaNetV 4 have purer Alpha attributes. The figure below shows the input data of AlphaNetV 4 and the regression target of AlphaNetV 4 .

Table 3-2 AlphaNet Input and Output

| div_open_close(t-4) | div_open_close(t-3) | div_open_close(t-2) | div_open_close(t-1) | div_open_close(t) |
|---|---|---|---|---|
| div_high_close(t-4) | div_high_close(t-3) | div_high_close(t-2) | div_high_close(t-1) | div_high_close(t) |
| div_low_close(t-4) | div_low_close(t-3) | div_low_close(t-2) | div_low_close(t-1) | div_low_close(t) |
| close(t-4) | close(t-3) | close(t-2) | close(t-1) | close(t) |
| div_vwap_close(t-4) | div_vwap_close(t-3) | div_vwap_close(t-2) | div_vwap_close(t-1) | div_vwap_close(t) |
| volume_sqrt(t-4) | volume_sqrt(t-3) | volume_sqrt(t-2) | volume_sqrt(t-1) | volume_sqrt(t) |
| return(t-4) | return(t-3) | return(t-2) | return(t-1) | return(t) |
| turnover(t-4) | turnover(t-3) | turnover(t-2) | turnover(t-1) | turnover(t) |

$R_{res(t+10)}$ 某股票未来10天经过Barra风格因子剥离的收益率

c. **Feature extraction**

i. **Feature extraction layer**

The feature extraction layer is the most critical component of AlphaNet, which draws on the idea of feature construction in genetic programming and convolutional network design in CNN. For traditional CNN, the local perception operation of the convolution kernel is related to the arrangement of data. This fixed arrangement does not exist in pure numerical data, which makes the local perception feature meaningless in pure numerical data. Therefore, AlphaNet uses multiple function operators as custom network layers to replace the original random convolution kernels for feature extraction . The following figure shows the feature extraction layer currently used in AlphaNet V 4. This chapter will analyze the representative custom network layers in detail .

Table 3-3 AlphaNetV4 factor equation

| name | definition |
|---|---|
| ts_corr(X,Y,d) | Pearson correlation coefficient of the time series values of X and Y over the past d days. |
| ts_co v ( X,Y,d) | The covariance of the time series values of X and Y over the past d days. |
| ts_stddev (X,d) | The standard deviation of the time series consisting of the X values in the past d days. |
| ts_zscore (X,d) | The mean of the time series consisting of the X values over the past d days divided by the standard deviation. |
| ts_return (X,d) | (X - delay(X, d))/delay(X, d)-1, where delay(X, d) is the value of X d days ago. |
| ts_decaylinear(X, d) | of the time series values composed of the X values of the past d days, with weights d, d - 1, ... , 1 (the sum of the weights should be 1, and normalization is required ), where the closer the day is to the present, the larger the weight. |

1. ts_corr ( X,Y, d )

    Assuming d = 3, the following figure shows the working mechanism of the ts_corr (X,Y,d) network layer. ts_corr ( X,Y,d) will traverse the two-dimensional data in the time dimension and the feature dimension. Similar to CNN, the stride of the step size is actually an adjustable hyperparameter. If stride = 1, the next calculation will move one step to the right in the time dimension. The calculation in the feature dimension reflects the difference from the CNN convolution layer. The CNN convolution operation can only perceive locally, but ts_corr ( X,Y,d) will traverse all types of data, and its calculation area does not have to be adjacent. For example, in Figure 8, it will traverse $C_9^2 = 36$ times. This avoids the problem of data arrangement caused by local perception in CNN, and can fully extract features. The result of the operation of ts_corr ( X,Y,d) is a two-dimensional "feature picture". For this picture, it can be directly flattened and input into the MLP, or it can continue to extract features or perform pooling on this basis. If you continue to extract features, you can nest operators, such as:

    $$ts\_corr ( ts\_corr (X,Y,3), ts\_corr ( Z , W ,3),3)$$

    Similar to ts_corr ( X,Y,d) is ts_cov ( X ,Y,d) .

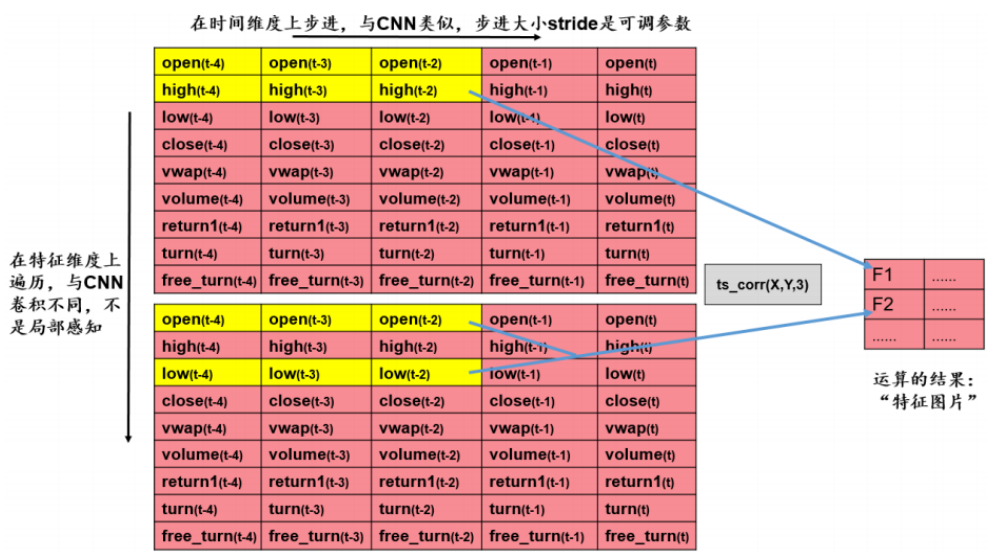

Figure 3-3 AlphaNet feature extraction layer structure 1

2. ts_stddev(X,d)

Assume d = 3. The following figure shows the operation mechanism of ts_stddev(X,3), which is similar to the convolution in CNN 1 * 3. The network layers such as ts_zscore(X,d) and ts_return(X,d) have the same operation mechanism as this network layer.

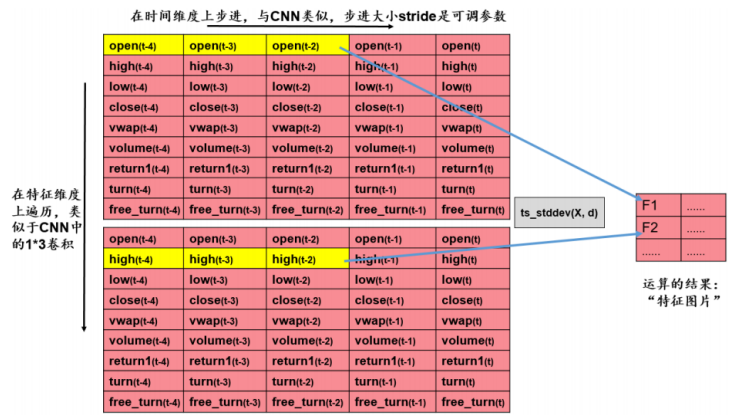

Figure 3-4 AlphaNet feature extraction layer structure 2

In this paper, in order to ensure that the dimensions of each feature matrix are the same, the input matrix of this paper is a 1 3 * k matrix, where k is not fixed and changes with the required feature matrix. This paper uses d = 2, 5, 10, 22 for feature extraction, and the output matrix is a 20 * 208 matrix (where 2 0 is the fixed sequence length in this paper. 208 = 2 * 78 + 4 * 13, there are 13 inputs in this paper, so the computational complexity of ts_corr and ts_cov is $C_{13}^2$= 78, and the computational complexity of the other four feature extraction functions is 13. ). So after passing through the feature extraction layer, a 4 * 20 * 208 tensor will be obtained.

ii. BN layer

In order for machine learning to be effective, the prerequisite that the data must have the same distribution must be met. As the number of layers of the neural network gradually increases, the data distribution will gradually shift and change due to changes in nonlinear activation functions. Some activation functions (such as Sigmoid and Tanh) may even cause the gradient of the low-level neural network to disappear, slowing down training convergence. Normalization is to standardize the distribution of the data input values of each layer of the neural network to a standard normal distribution with a mean of 0 and a variance of 1 through certain normalization methods, so that the activation input value falls at a position with a larger gradient of the activation function, thus avoiding the problem of gradient vanishing. Moreover, the larger the gradient, the faster the learning convergence speed, which greatly speeds up the training speed of the model [19].

$Z^l$ Assume that the calculation result of the lth layer of the neural network, m is the number of samples in each batch, and we have:

The mean of samples in each batch at layer l:

$$\mu = \frac{1}{m}\sum_{i=1}^{m} Z^{l(i)}$$

Formula 3-4

The variance of samples in each batch at layer l:

$$\sigma^2 = \frac{1}{m}\sum_{i=1}^{m}(Z^{l(i)}-\mu)^2$$

Formula 3-5

The result of each BN layer is:

$$\hat{Z}^l = \gamma * \frac{Z^l - \mu}{\sqrt{\sigma^2 + \varepsilon}} + \beta \qquad \text{Formula 3-6}$$

In the above formula, in order to enhance the expression ability of the BN layer, two optimizable parameters $\beta$ and are introduced $\gamma$. Without these two parameters, the BN layer is actually z-score standardized, so the BN layer has more plasticity and operation space than the usual standardization. The figure below shows the result of the BN layer. From the feature distribution histogram, before the BN layer is standardized, the value range of the feature extracted by the ts_corr layer is ( -0.7,0.4) , and the value range of the feature extracted by the ts_std layer is ( 0,23000000) , which is a big difference. After the BN layer is standardized, the value range of the features is very close, all in ( -1.5,2) , and the dimensions between the features are comparable.

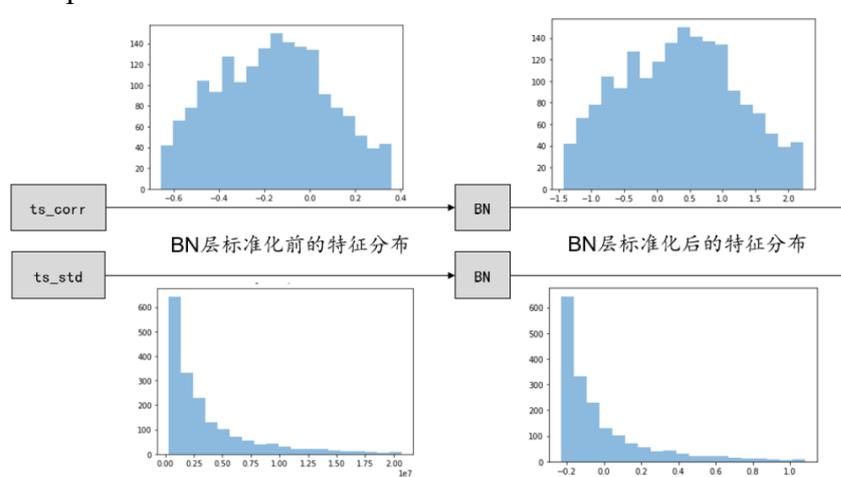

Figure 3-5 Comparison of results before and after standardization

The BN layer only standardizes the data and does not change the size and shape of the matrix, so after passing the BN layer, the size of the matrix is still 4* 20 * 208 .

iii. **Spearman correlation coefficient DropOut layer**

In machine learning, the DropOut layer is usually used to control overfitting and improve the model training speed . However, in traditional convolutional neural networks, the DropOut of the feature extraction layer is random [18] . This paper finds that random DropOut may make the model unstable. Therefore, this paper borrows the idea of Dropout and uses the calculation of the Spearman correlation number to implement a custom Dropout mechanism for the feature extraction layer of AlphaNet : After the first batch of feature extraction is completed, the function operators (ts_cor r , ts_co v, etc.) are sequentially traversed , and the features with correlation coefficients greater than the critical value (the critical value in this paper is 0.8 ) are discarded, and the features less than the critical value are retained, and the function operators of the retained features are remembered. When performing feature extraction in the subsequent batch, only the retained function operators are extracted. This mechanism can improve the model from the following three aspects:

First, it can save the computational overhead of operator functions and improve training speed. In addition, due to the reduction in computational overhead, more original features can be input into the model, which may improve the model's predictive ability.

Second, compared with sampling DropOut, although this traversal method has the complexity of correlation coefficient calculation, it avoids random sampling and losing important features, which reduces the correlation between features in the input layer network and controls overfitting to a great extent .

Third, this method enhances the interpretability of the model and facilitates model reuse . In the test, AlphaNet with the Spearman correlation coefficient Dropout mechanism showed a slight improvement in performance in terms of revenue, while significantly reducing training time.

In the original feature extraction layer, this paper has a total of 13 inputs, so the computational complexity of ts_corr and ts_cov is $C_{13}^2$= 78 , and the computational complexity of the other four feature extraction functions is 13 , so the overall complexity is 2* 78 + 4 * 13 = 208 . However, after DropOut of the Spearman correlation coefficient, this paper only retains 103 factors with low correlation coefficients, and the reduced factors do not affect the model's fitting ability.

After Spearman DropOut , this paper will get a 4 *20*103 tensor, in which the 20*103 "data image" generated by d= 10 is used as the "primary data image" in this paper, and the "data images" of d= 2 , d= 5 and d= 20 are used as "auxiliary data images". The extraction stride of these "data images" is 1, and in order to facilitate the subsequent residual optimization operation, the time series length of each "data image" is 20. The arrangement is shown in the figure below.

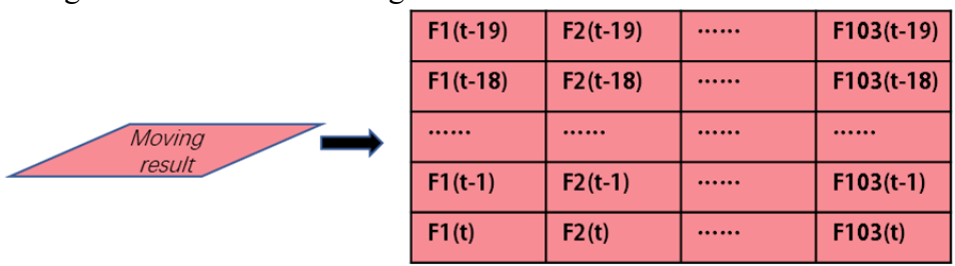

Figure 3-6 AlphaNetV 4 data image

### d. Neural network layer construction and prediction

#### i. Bi-LSTM Layer

Bi-LSTM stands for Bi-directional Long Short-Term Memory. It is a time series representation method based on a bidirectional LSTM, which is composed of a forward LSTM and a backward LSTM. Bi-LSTM can better handle the problem of gradient disappearance and explosion, and can obtain relevant long-term information in the stock sequence. Since LSTM is a combination of bidirectional LSTM, after inputting a 20 *103 data image, a 20 * 206 matrix is obtained , in which the second dimension contains both the information of the 103 features extracted forward and the information of the 103 features extracted backward .

#### ii. Transformer residual optimization

, the "primary data image" extracted from d = 10 is input into Bi-LSTM to obtain an initial matrix of 20 * 206. The "auxiliary data images" extracted from d = 2 , d = 5 and d = 20 are respectively passed through two layers of stacked Transformer Encoder, and obtain three matrices with the same dimensions as the Bi-LSTM result (2 0*206 ). The reason for using Transformer to extract instead of directly copying the "auxiliary data image" forward and backward is: (1) increasing the complexity of the model, so that the model has a better fitting effect under large amounts of data. (2) converting the residual items on each subsequent data point into a collection of time series data, rather than the information on the data point at that time, as shown in the figure below.

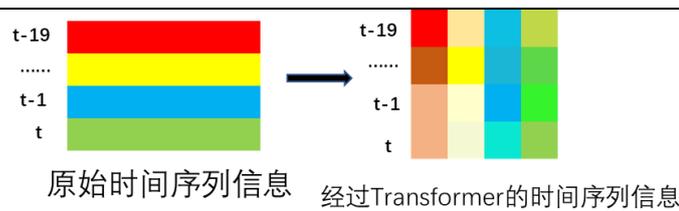

Figure 3-7 Transformer feature extraction

This paper adds the two matrices extracted by Transformer to the initial matrix and performs Batch Normalization, this step is called residual optimization. There are two problems in deep learning that will affect the training accuracy of the model. The first is the problem of gradient explosion and gradient diffusion. This problem was previously discussed in Batch Normalization and LSTM can be effectively solved. Another problem is the problem of network degradation. According to common sense, if there is a K-layer network f (x) that is the current optimal network, then a deeper network can be constructed, and its last few layers are just the identity mapping of the K-th layer output of the network f (x) . This should achieve the same result as f (x) ; perhaps K is not the so-called "optimal number of layers", so a deeper network can achieve better results. In short, compared with a shallow network, a deeper network should not perform worse. But the reality is often that, under the premise of non-overfitting, as the network depth increases, the network performance usually gradually increases to saturation and then decreases rapidly. A reasonable guess for this problem is that the identity mapping is not easy to fit in a neural network. To solve this problem, ResNet [12] proposed in 2015 constructs a natural identity mapping by residual means. Assuming that the input and output dimensions of the nonlinear unit of the neural network are consistent, the function f (x) to be fitted in the neural network unit can be split into two parts, namely:

$$f^{(x)} = F(a^{x-1}) = a^{x-1} + R(a^{x-1}) \qquad \text{Formula 3-7}$$

Here R (x) is the residual network. At the high level of the network, learning an identity mapping f (x) -> a(x-1) is equivalent to making the residual R (x) tend to 0. This paper wants to predict the price in the next 10 days. If we only enter the data image obtained by d = 10 into the network through LSTM, it may be difficult to fit, because the change in the yield in the next 10 days may not only be related to the result of the moving feature extraction function obtained by d = 10 , but also to the moving feature extraction function of d = 2 , d = 5 , d = 20 , d = 30 , and even longer d. The most classic theory in finance is the concept of "golden cross" and "dead cross": when the short-term moving average breaks through the long-term moving average from bottom to top, it is a buy signal; if the short-term moving average falls through the long-term moving average from top to bottom, it is a sell signal [19] . The concepts of golden cross and dead cross exist in the long-term and short-term moving averages of closing prices, but in various complex features, there may be similar or more complex signals, which can also be compared to the difference between short-term investors and medium-term and long-term investors in the financial market. Therefore, this article combines this feature of the financial market with the residual network training method, trains the Transformer Encoder through the residual method, and finally integrates the information of four data images to obtain a 20 * 206 matrix .

iii. Attention Mechanism

Finally, this paper performs Attention weighted averaging on the final 20 *206 matrix according to the state of each hidden layer h obtained by the first Bi-LSTM, and obtains the final real number output, which is fitted with the future 10 - day yield set in this paper.

### e. AlphaNetV4 model training and construction details

#### i. Increase the weight of multi-head samples

Since there are limited means of shorting A-shares and it is difficult to obtain short returns, a common problem in factor investing in A-shares is how to discover factors with significant long returns. In AlphaNet, the model can be guided to discover factors with significant long returns by increasing the weight of long samples.

As shown in the following formula, this paper considers the loss function from the perspective of a single sample. Let the actual rate of return of sample i be $y_i$, the $w_i$ rate of return predicted by the model be $y(x_i)$, $L(y_i, y(x_i))$ be the loss function of sample i, and be the weight of the loss function of sample i. Then the Loss of the overall loss function is the weighted sum of the loss functions of all samples. By default, all samples have $w_i$ equal weights. For AlphaNet, this paper guides the model to mine the weights of factors with significant long-term returns by increasing the weights of long-term samples in the training set, so that the model focuses more on optimizing the prediction effects of these samples.

$$\text{Loss} = \sum_{i=1}^{n} w_i * L(y_i, y(x_i)) \quad 其中 \quad w_i = \frac{y_i}{\sum_{i=1}^{n} y_i} \qquad \text{Formula 3-8}$$

#### ii. Sharp EarlyStopping

In the field of deep learning, the training cycle is a critical issue. Too few training rounds will lead to underfitting, while too many training rounds will lead to overfitting and waste of computing resources. Early Stopping is used to help this paper solve this problem. Its function is to stop training when the performance of the model on the validation set no longer increases, thereby achieving the effect of full training and avoiding overfitting. Since this paper applies deep learning technology to the stock market, in the selection of training cycles, this paper does not use the traditional MSE to set Early Stopping. Rational investors will only choose two investment portfolios: the first is the investment portfolio that maximizes the expected return under a given risk; the second is the investment portfolio that minimizes the risk under a given expected return [21]. Therefore, in model training, this paper sets the training cycle through the backtested Sharpe ratio. When the Sharpe ratio does not increase for 10 consecutive cycles, this paper will interrupt the model training and select the model with the highest Sharpe ratio as the best model. The purpose of this is to select the optimal model based on the actual stock market, rather than defining the optimal model parameters based on the loss size of the training set.

### iii. AlphaNetV4 model construction details

Table 3-4 AlphaNetV4 model construction details

| Network composition | Included Components | Parameters and Description |
|---|---|---|
| Feature extraction layer+ Spearman DropOut | ts_corr(X,Y,d)<br>ts_cov(X,Y,d)<br>ts_stddev(X,d)<br>ts_zscore(X,d)<br>ts_return(X,d)<br>ts_decaylinear(X, d)<br>BN<br><br>Spearman correlation coefficient DropOut | 1. In the custom network layer, stride = 1, the time series length is 20.<br>2. Each custom network layer is followed by a BN layer<br>3. There are three ways to extract features simultaneously in the feature extraction layer, namely d= 5, d= 10, and d= 20.<br>4. Only the first extraction of the first batch of d= 10 has the Spearman correlation coefficient judgment, and then all filtering feature extraction is performed based on the results of this judgment. The critical value of the Spearman correlation coefficient is 0.8. |
| Bi-LSTM | Bi-LSTM | 1. Only the feature matrix of d = 10 enters this network. The dimension parameters of LSTM unidirectional input and output are the same. Since it is bidirectional, the total output dimension is twice the input dimension.<br>2. The DropOut ratio is 0.2. |
| Transformer | Transformer | d = 5 and d = 20 enter the Transformer. In order to keep the dimensions consistent with Bi-LSTM, |

| | | each feature matrix must pass through 2 Transformers, so there are 4 Transformers in total. Encoder, and these four Transformer extraction layers do not share parameters. |
|---|---|---|
| LSTM Attention | Attention Mechanism | The state of each hidden layer of the first Bi-LSTM is weighted using Attention to obtain the final output. |
| Output Layer | 1 neuron | 1. Activation function: Leaky Relu. 2. Weight initialization method: truncated_normal. 3. The fitting target is to use the Barra model to eliminate the return of style factors. |
| Other model parameters | 1. Loss function: mean square error (MSE). 2. Optimizer and learning rate: Adam, Learning Rate Decay (from 1e-3 to 1e-7). 3. Training period: 90 periods with early use of Sharpe ratio stopping. 4. From January 1, 2019 to June 1, 2021, training and prediction are performed every 6 months, for a total of 5 training and predictions. The training sample data is from January 1, 2016 to the starting date to be predicted. | |

## 4. Comparative analysis of the prediction capabilities of AlphaNetV4 and existing models

In response to the problems existing in AlphaNetV1 to V3, this paper improves the fitting effect by 1. increasing the sequence length and reducing the step size of the extraction layer; 2. reducing the correlation of the original input; 3. using the Spearman correlation coefficient to design the DropOut layer to replace random sampling to enhance the stability of feature extraction; 4. using Bi-LSTM to replace the original double-layer LSTM and double-layer GRU design to enrich the design of the extraction layer, and enhancing the learning ability of the model through the residual optimization of Transformer, and proposes AlphaNetV4. In this chapter, the AlphaNetV4 proposed in this paper is empirically compared and analyzed with the original AlphaNet version in view of the above improvements.

### a. Input data to improve empirical analysis

#### i. Input data preparation

The data in this article comes from the Wind database. This article captures the data of all A-share stocks from January 1, 2016 to June 1, 2021, and excludes stocks such as ST, PT, daily limit, and daily limit that have no positive impact on the fitting results. In these more than 1,000 trading days, there are about 4,300 stocks traded every day, so the input data of this article is more than 4 million rows of panel data. This article predicts the stock price changes from January 1, 2019 to June 1, 2021, and re-forecasts every 6 months. Each re-forecast uses the data from January 1, 2016 to the day before the prediction as the in-sample data for training, and a total of 5 cycles are predicted.

#### ii. Rationality analysis of reducing sequence length

AlphaNetV2 and V3 abandoned the structure of the pooling layer based on AlphaNetV1, and improved it to LSTM and GRU for time series fitting. The advantage of LSTM is that it can solve the problem of long-term dependence, but in previous versions, the step size of feature extraction used by the author was very large, with a step size of 10 in AlphaNetV1 and a step size of 5 in AlphaNetV2. This resulted in a very small number of time series obtained by the feature extraction layer under the same input matrix, only 3 in AlphaNetV1 and only 5 in AlphaNet. Through experiments, this paper found that under the same input matrix, reducing the step size and increasing the length of the time series has a very positive effect on the final regression of the model. The results are shown in the figure below, where the best effect is obtained when Timestep = 2 and 1, so this paper reduces the step size of the feature extraction layer to 1, and because the Transformer residual structure after this paper must ensure the consistency of the number of dimensions of all feature extraction layers, this paper fixes the length of the time series to 20 for feature extraction.

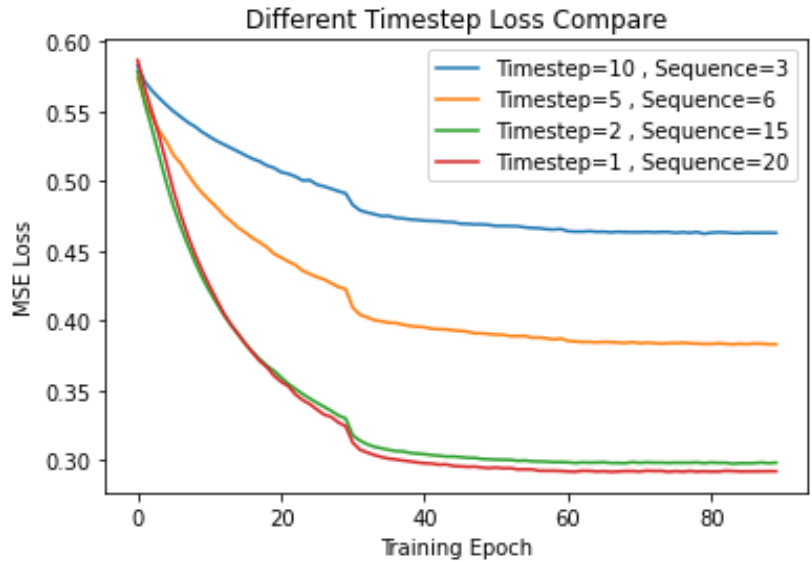

Figure 4-1 Loss comparison of different time series lengths

this paper fixes the sequence length to 20 is that the longer the time series length, the slower the training speed will be. LSTM cannot be run in parallel. When the step size becomes longer, the number of forward and backward propagation steps also increases, and the time complexity is higher than the linear growth. Even if LSTM has the ability to avoid long-term forgetting, it still has an upper limit. The hidden layer information capacity of LSTM always has a saturation point. When the sequence length continues to grow, its final fitting effect tends to converge.

iii. Rationality Analysis of Feature Input Optimization

The features of this article are improvements made on AlphaNetV 1 and V2, as shown in the figure below.

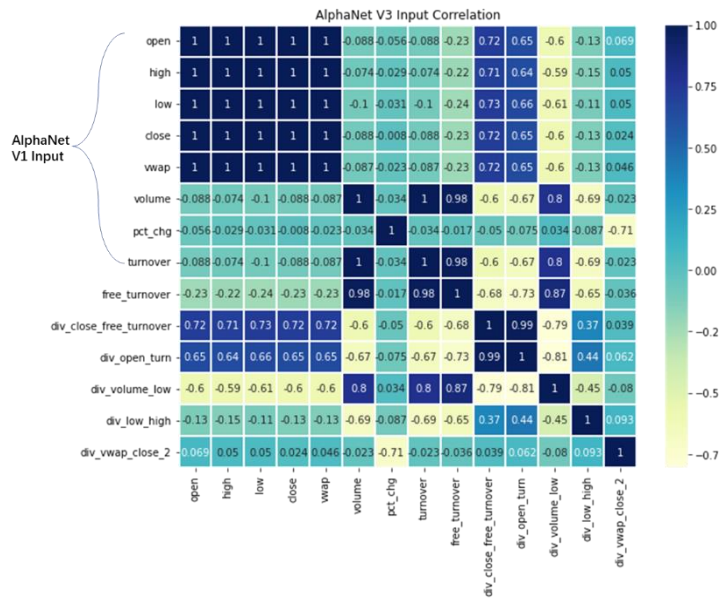

Figure 4-2 AlphaNet feature correlation coefficient matrix

In the original AlphaNetV 1 , there are only 9 inputs, and through the correlation coefficient matrix, 5 types of price information are found. The input correlation between turnover, free_turnover , and volume is very high, which will have a very serious impact on the diversity of feature extraction in the following article. It not only increases the computational overhead of this article, but also does not bring about the gain of effective information. From AlphaNetV 1 to V 2 , the original author considered the effectiveness of ratio features and expanded 6 ratio features [16], but there are also features with high correlation between these 6 features (for example, div_close_free_turnover and div_open_turn and div_vwap_high and div_low_high ) . Therefore, in the AlphaNetV 4 proposed in this article , some adjustments are made to some inputs and outputs:

(1). First, for the five types of price information, this article retains the closing price close, and the remaining four prices are presented in the form of change rate. For example, div_open_close refers to the opening price of a stock on that day minus the closing price divided by the closing price.
(2). Since the correlation between volume and turnover is high, this article raises the volume to the eighth power to minimize the linear correlation between the two variables. This also helps to reduce the instability of normalization caused by the excessive dimension of volume. The correlation between turnover and free_turnover is too high, so this article deletes the variable free_turnover .
(3). Since close and open are highly correlated, and free_turnover and turnover are highly correlated, the correlation between the ratio factors close/free_turnover and open / turnover calculated by AlphaNetV 2 is also very high. Therefore, this paper divides open / turnover by close/free_turnover while retaining close/free_turnover to generate a new variable div_price_turnover to replace the original open / turnover.
(4). Similarly, since the correlation between vwap and low is very high, the correlation between the ratio factors vwap / high and low / high is particularly high. In addition, this article also takes this type of information with growth rate into consideration when processing price information in the first step. Therefore, this article eliminates vwap / high and retains low / high.

The final correlation matrix of the original input and the loss optimization diagram of the improved method compared with the original method are shown below.

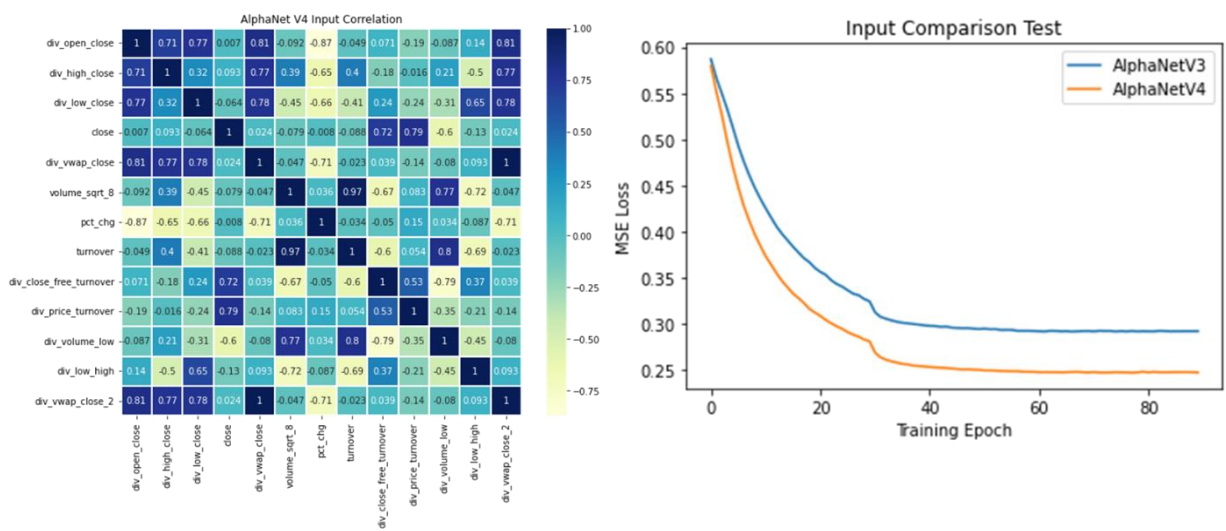

Figure 4-3 AlphaNetV 4 optimized feature correlation coefficient matrix and loss comparison

b.      Reasonable analysis of Spearman DropOut layer

Spearman DropOut is an improvement on the AlphaNetV 3 sampling traversal DropOut layer. In " AlphaNet Improvement: Structure and Loss Function - Huatai Artificial Intelligence Series No. 46 " [18] released by Huatai Securities on July 6, 2021 , the author used the sampling traversal method to control overfitting and improve the model training speed. The sampling traversal method is as follows. Assuming that this article has 9 features and a total of 5 days of data, the ts_corr (X,Y,d) function needs to calculate the correlation coefficient of the two time series X and Y within d days, which requires convenient calculation $C_9^2$ = 36 times, and the computational complexity is O ( $n^2$ ) . Sampling traversal randomly samples n ( 0<n<9) features from the 9 original features and traverses them two by two.

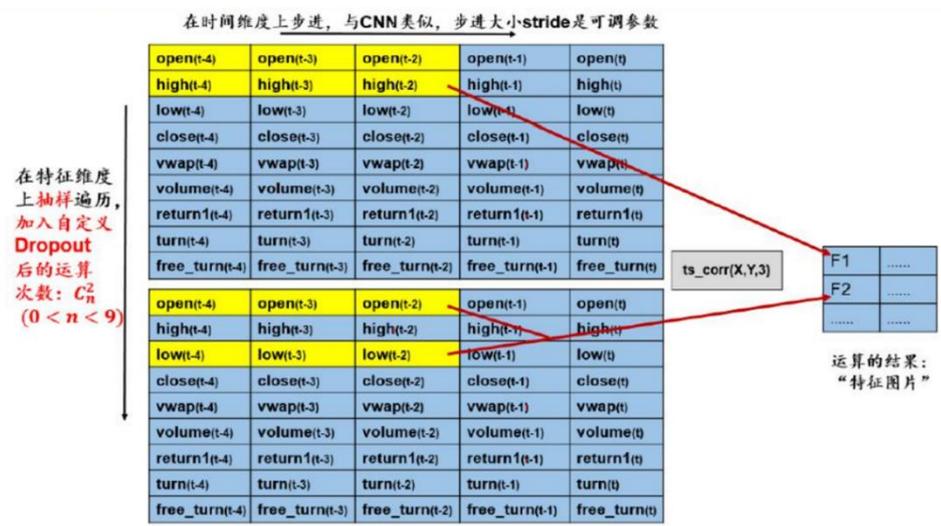

Figure 4-4 AlphaNetV 3 sampling traversal

This paper believes that sampling traversal has the following disadvantages:

First, if the sampling traversal is performed inside the neural network, the different features selected each time will affect the parameter update of the subsequent network, resulting in the gradient being unable to decrease and fitting being impossible.

Second, if the sampling traversal is placed before the neural network, the network trained by the word sampling traversal may have a poor fitting effect because the sampling traversal ignores more important factors. This paper also conducts an empirical analysis of this phenomenon, as shown in the figure below. The loss of the network trained by each sampling will be higher than that of the network trained without sampling, and the loss in the best case is also convergent to the original network. Therefore, if the sampling traversal is used for word training, the effect of sampling traversal is extremely unstable.

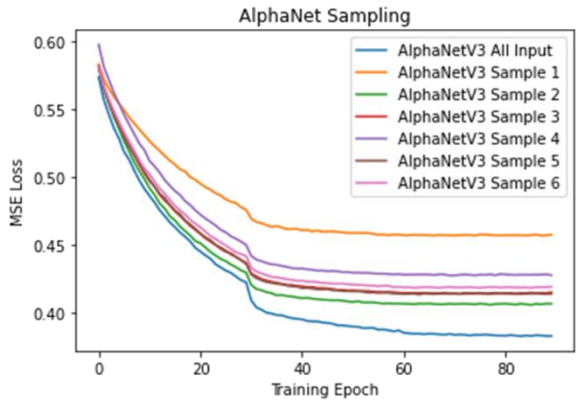

Figure 4-5 Loss comparison of AlphaNet before and after sampling

Third, in AlphaNetV 3, the author also considered that sampling traversal would make the network unstable, so the author adopted the method of integrating 10 models for fitting. This paper believes that the method of integrated learning can only make the results of the final integrated model converge to the results of the model trained by all the features of the original input, and integrated learning violates the purpose of sampling traversal to improve the model training speed.

In order to truly control overfitting, improve training speed, and improve the efficiency of single network training by non-integrated methods, this paper replaces random sampling traversal with Spearman correlation coefficient traversal, and discards (DropOut) the extracted features with correlation coefficients greater than the set critical value (the critical value set in this paper is 0.8 ) . Using the Spearman correlation coefficient filtering method, the traversal inside the neural network makes the discarded variables fixed and will not cause problems in updating the parameters of the later network layers. And after the feature reduction, the sampling and operation time of the model can be reduced to 70% of the original , and the performance of the model can be maintained. The training effect is shown in the figure below.

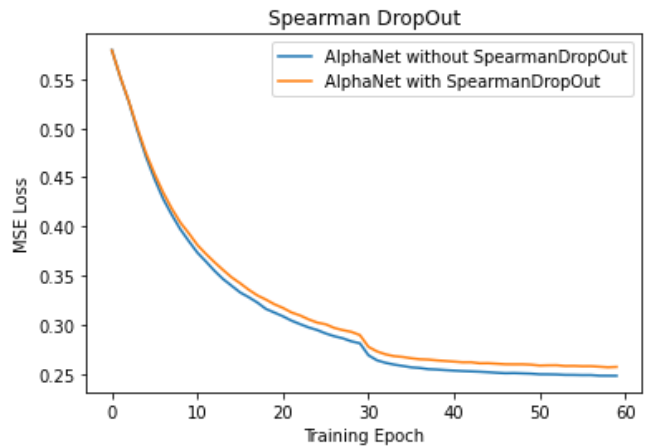

Figure 4-6 Spearman Loss comparison between DropOut and full sample prediction

's suggestion that sampling training in AlphaNetV 2 is conducive to controlling overfitting, so this paper incorporates this sampling DropOut mechanism into LSTM. This paper randomly and temporarily deletes 20 % of neurons and their related connections during each round of training, thereby accelerating the training process and achieving better training results.

Theoretically, the training can be further accelerated by omitting the calculations of neurons and synapses that are temporarily deleted in Dropout. However, this paper separates the feature extraction layer and the network for calculation, so this method is not tried.

c. **Rationality analysis of network structure optimization**

The Transformer residual design in this paper is based on the improvement of the two-layer feature extraction in " Revisiting AlphaNet: Structure and Feature Optimization " [16] released by Huatai Securities on August 24 , 2020. In " Revisiting AlphaNet: Structure and Feature Optimization ", the author used different lookback intervals ( d = 10 and d = 5) , merged this feature extraction layer after passing through GRU, and then fitted it through MLP. The structure of the model is shown in the figure below.

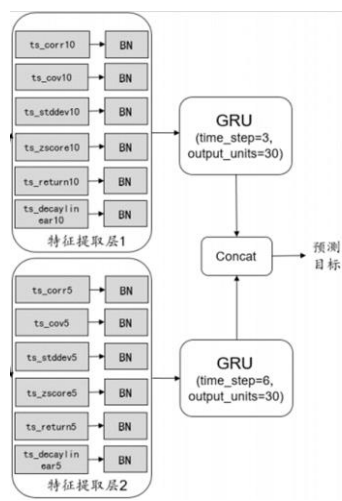

Figure 4-7 GRU structure of AlphaNetV 2

This paper found that the problem with the original Concat method is that the correlation between the same variables in feature extraction layers 1 and 2 is particularly high. The method of extracting separately using d= 10 and d= 5 and then concat cannot effectively extract the influence of the features of d= 5 on the features of d= 10 . In short, this method cannot provide effective support for model fitting. As shown in the figure below, the distribution of the mean of the parameter matrix obtained by d= 10 and d= 5 is almost the same, which shows that the two parameter matrices are actually fitting in the same direction. This paper found that the parameter matrix of d= 10 is darker, which shows that most of the significance is reflected in the parameter matrix of d= 10 .

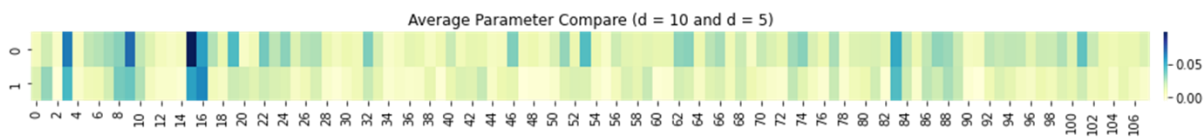

Figure 4-8 GRU fitting parameter matrix

This paper can also draw the same conclusion from the results of model training, as shown in the figure below. For the regression with the next 10- day return as the target, d= 5 has the worst performance, d= 10 has the best performance, and the effect of merging the feature extraction layers of d= 5 and d= 10 is not as good as the effect of the feature extraction layer obtained by d= 10 alone.

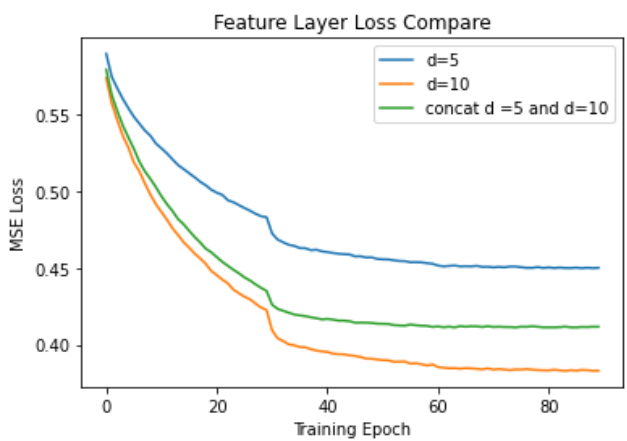

Figure 4-9 Comparison of fitting losses extracted at different time lengths

Such a result does not mean that the "weekly average" feature matrix obtained by d = 5 is not effective for the final result. In the stock market, the concept of entering the market with a golden cross and leaving the market with a dead cross is a frequently used technical analysis. So how to increase the impact of the d = 5 "weekly average" feature matrix on regression is also the problem that this paper wants to solve. This paper uses the Transformer + residual structure. On the basis of retaining the d = 5 feature matrix, it adds a longer extraction sequence d = 22 and a shorter sequence d = 2 matrix to represent the daily average and monthly average of the features. These two feature matrices are fused by Transformer, added to the original matrix by residual, and the matrix parameters are updated by learning the residual. This fine-tuning method brings the features extracted by d = 2 , d = 5 and d = 22 into the main matrix, achieving a better fitting effect. The effect comparison is shown in the figure below.

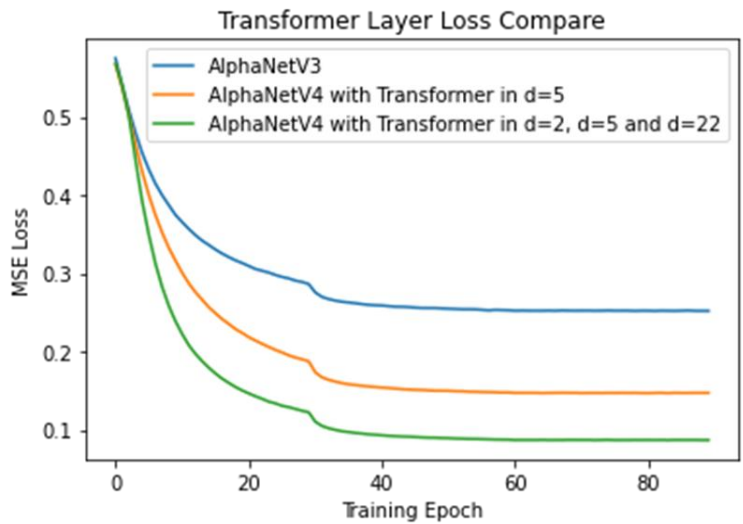

Figure 4-10 Loss comparison between AlphaNetV 4 with Transformer and previous models

After residual learning of the Transformer layer, this paper uses the attention weighting of each hidden layer state of Bi-LSTM to obtain the final output. The Bi-LSTM+Attention mechanism increases the speed of model convergence compared to using only LSTM, as shown in the following figure.

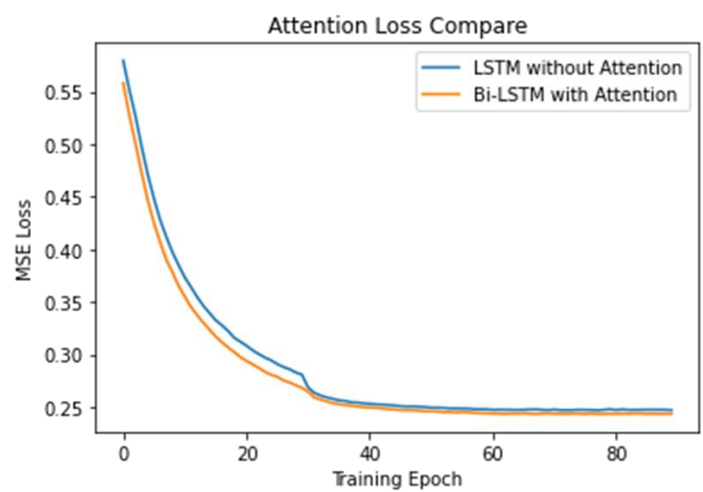

Figure 4-11 Loss comparison between AlphaNetV 4 with Attention and previous models

d. **AlphaNetV 4 Empirical Analysis Summary**

the problems of AlphaNetV 3 , the AlphaNetV 4 proposed in this paper has been optimized from the aspects of sequence length, original input, DropOut layer design, network structure, etc. Among them, by changing the sequence length from 5* 6 to 20 * 1 , the Loss value is reduced from 0.4 to 0.3 ; by redesigning the original data input, the Loss is further reduced from 0.3 to 0.2 ; by designing the Spearman DropOut layer, the model training time is reduced to 70 % of the original time, and there is almost no impact on the accuracy; by designing the Transformer, Bi-LSTM and Attention layers, the Loss is further reduced to below 0.1 . In general, this paper has reduced the AlphaNet Loss from the original 0.4 to below 0.1 through optimization in these aspects , and reduced the computational overhead of the model by 30 % , achieving a huge optimization.

**5. based on AlphaNet V4**

This chapter will test the improved AlphaNetV 4 based on stock volume and price data . The rebalancing cycle of this article is 10 days , and the model interpretability analysis is performed.

   a. **Single Factor IC Test**

This paper regards the prediction results of AlphaNetV 4 on each interface as a synthetic single factor and conducts a single factor IC test. The backtest period is from January 1, 2019 to June 1, 2021, and calculations are performed every 10 trading days .

The figure below shows the test results of IC. In the AlphaNetV 4 model, the average IC reached 8.08 % and the IR value was 1.52 , with significant incremental information.

Table 1-1 AlphaNetV4 single factor IC test results

| Model | IC mean | IC standard deviation | IR value |
| --- | --- | --- | --- |
| AlphaNetV 1 | 6.00 % | 5.55 % | 1.08 |
| AlphaNetV 3 | 6.68 % | 5.96 % | 1 .12 |
| AlphaNetV 4 | **8.08%** | **5.3 %** | **1.52** |

   b. **One-factor hierarchical test**

This paper regards the prediction results of AlphaNetV4 as a synthetic single factor and conducts a single factor hierarchical test. The test method is as follows:
1. The number of layers for the delamination test is 5.
2. The data used in the test is consistent with the IC test.
3. Position switching: The stratification is obtained through the prediction results, and the stratification results are used to switch positions according to the vwap price on the next trading day. The transaction fee is set at 0.002 % and the stamp duty is 0.015 % .
4. Layering method: Sort all the stocks in the stock pool from large to small according to the predicted factor values, divide them into N equal layers, and assign equal weights to the stocks in each layer.

Table 5-2 AlphaNetV4 single factor stratification test results

| Model | Annualized excess returns of tiered portfolios 1 to 5 (from left to right) | | | | |
| --- | --- | --- | --- | --- | --- |
| AlphaNetV 1 | 1 3.07 % | 5.48 % | 4.38 % | -1.10 % | **-12.70 %** |
| AlphaNetV 3 | 1 7.08 % | 1 1.22 % | 6.74 % | 0.00 % | -19.89 % |
| AlphaNetV | **1 7.42 %** | **1 1.62 %** | **8.09 %** | **0.09 %** | -20.26 % |

|   | 4 |   |   |   |   |   |

The figure above shows the results of the stratified tests of each model . When the transaction cost is 0.02% , after AlphaNetV 4 is neutralized by five factors, the annualized excess return of the Top group reaches 17.42 % . The stratification effect is more obvious than the previous model, and the excess return is higher. The figure below is the stratified excess return change chart of AlphaNetV 4 in the complete backtest period.

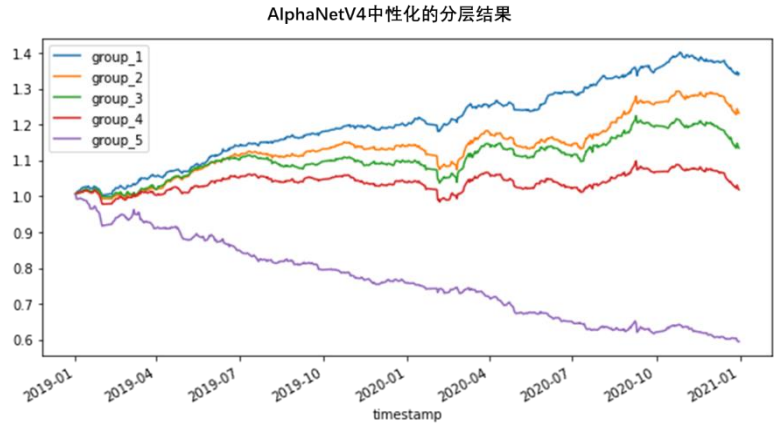

Figure 5-1 AlphaNetV4 layered test results

### c. Backtesting using the CSI 500 Enhanced Strategy

This paper regards the prediction results of AlphaNetV 4 on each section as a synthetic single factor, constructs an industry- and market-value-neutral A-share stock selection strategy relative to the CSI 500, and backtests it. The test method is as follows:
1. The data used in the test is consistent with the IC test and layered test.
2. Position switching: The stratification is obtained through the prediction results, and the stratification results are used to switch positions at the vwap price on the next trading day. The transaction fee is set at 0.002 % and the stamp duty is 0.015 % . The turnover rate of each position switching is limited to less than 30 % .

Table 5-3 AlphaNetV4 backtest results

| Model | time | Yield | Number of shares | Standard Deviation of Return | Sharpe Ratio | Maximum Drawdown | Turnover rate |
|---|---|---|---|---|---|---|---|
| AlphaNetV 1 | 2019 | 3.82% | 209 | 6.24% | 0.64 | 3.89 % | **8.29%** |
|  | 2020 | 7.83% | 2 34 | **6.45%** | 1.24 | 3.46 % | **7.45%** |
|  | First half of 2021 | -0.46% | 2 49 | 11.17% | - 0.11 | 9.46% | **7.93%** |
| AlphaNetV 3 | 2019 | 12.70% | 2 11 | 6.38% | 1.98 | 3.73% | 14.41% |
|  | 2020 | 9.78% | 2 24 | 6.79% | 1.47 | 4.50% | 13.54% |
|  | First half of 2021 | 2.17% | 2 53 | **11.08%** | 0.53 | **6.74%** | 14.24% |
| AlphaNetV | 2019 | **22.49%** | 2 10 | **6.08%** | **3.53** | **3.05%** | 14.56% |
| 4 | 2020 | **22.07%** | 2 23 | 6.64% | **3.26** | **3.88%** | 15.57% |
| | First half of 2021 | **2.76%** | 2 53 | 11.87% | **0.61** | 7.06% | 14.86% |

The above table shows the results of the backtest. In 2019 and 2020 , the excess returns of the improved AlphaNetV 4 in this article exceeded 20 % ; during the period from 2019 to the first half of 2021 , the Sharpe value of AlphaNetV 4 was also higher than that of the previous two models, and the drawdown was more stable.

### d. Model interpretability analysis

This section uses the absolute value of the neural network parameters to analyze the interpretability of the AlphaNetV4 model. As shown in the figure below, this paper finds that most of the factors with significant parameters are composed of composite factors. These significant factors can also provide a certain basis for future factor mining, and can also reduce the dimension of factors based on the ranking of factor significance.

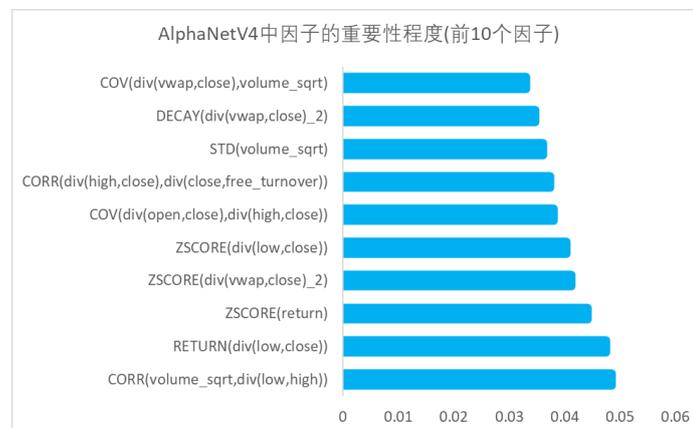

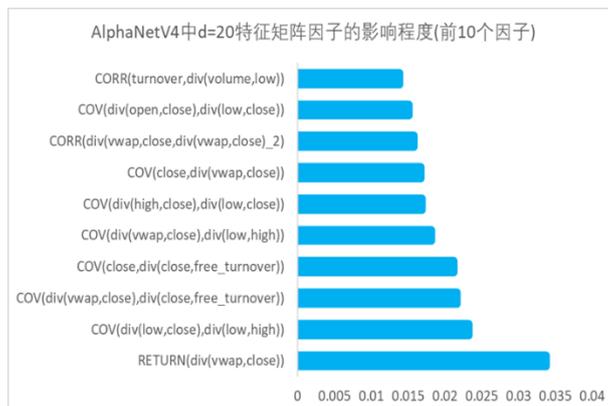
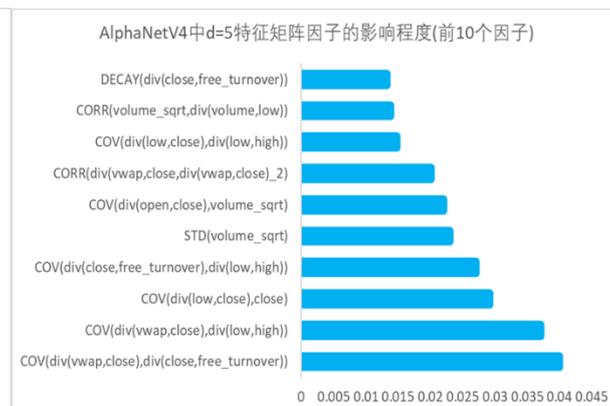

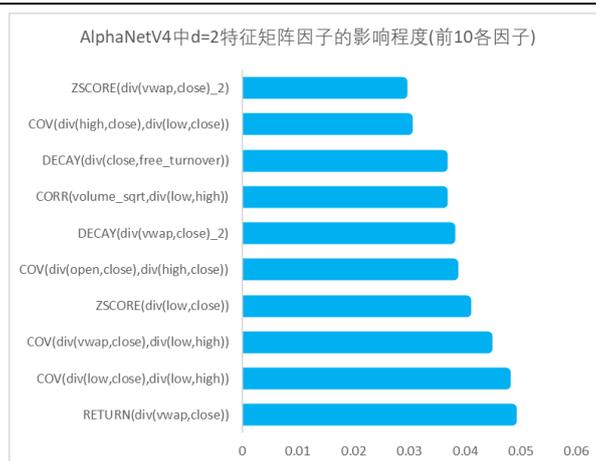

Figure 5-2 AlphaNetV4 model parameter significance

e. **Backtesting Summary**

compares the AlphaNetV 4 of this paper with the original model from the perspective of IC test, stratification test, and backtest based on the CSI 500 Index . Overall, AlphaNetV 4 has a better information ratio, is more obvious in stock stratification, and has an annualized excess return of more than 7%-10% compared to the original model . In addition, this paper provides factor mining directions for future research by visualizing the significance of variables.

6. **Summarize**
   1. deep learning makes it possible to mine factors with neural networks. Unlike traditional machine learning, deep learning models have a large number of parameters, can model spatiotemporal locations, and capture information features of high-dimensional data such as images, voice, and text. Their performance in large data sets and complex scenarios can beat traditional machine learning models (such as XGBoost, random forest, etc.), and neural networks can learn features, and achieve end-to-end. Through end-to-end methods, features can be directly extracted from raw data and predictions can be made for the target, avoiding information interference and loss caused by human information processing and extraction. Neural networks are algorithms that can run on GPUs and can achieve the desired fitting effect faster. CNN, which is commonly used in image recognition, RNN for time series analysis, and Transformer for speech recognition in deep learning are all shining in their respective fields. To achieve excellent end-to-end learning results, it is necessary to combine the flexibility of neural networks and the domain knowledge of the problem to be solved, and design the network in a targeted manner. This article uses the craze of artificial intelligence to apply these technologies to the factor mining and synthesis problems in the field of financial quantification, reconstruct the original AlphaNet, and design AlphaNetV 4 .
   2. This paper introduces the structure and technical details of the improved version of AlphaNetV 4. AlphaNetV 4 modifies some input features based on V 3 and uses Spearman The DropOut layer is designed to prevent overfitting and speed up model training. In terms of the feature extraction layer, AlphaNetV 4 inherits the feature extraction layer of the previous model. In terms of the regression layer, AlphaNet not only modifies the structure of the LSTM layer, but also performs residual learning through the Transformer results of different feature matrices to increase the model's fitting ability. In terms of the output layer, AlphaNetV 4 uses the Attention method for weighting to speed up the convergence of the model when fitting.
   3. Based on the quantitative and price data, this paper tests the performance of AlphaNet's stock

selection strategy under 10 - day rebalancing. Based on all the quantitative and price data of A-shares, this paper introduces the IC test comparison, stratified test comparison and excess return backtest comparison of AlphaNet and previous models based on CSI 500. The AlphaNetV 4 proposed in this paper has better performance in these tests. Finally, this paper also conducts an interpretability analysis of AlphaNetV 4 .

The application of deep learning algorithms in quantitative investment, in addition to studying the fluctuations of individual stock prices and finding excellent investment portfolios to find positive alpha, also has broad research prospects for optimizing certain investment processes to make investment steps automated: for example, how to solve the impact of large orders in the market, and the supervision of high-risk behaviors of fund managers. In addition , neural networks also have development potential in market stability control, such as how to predict and control the impact of investor sentiment , etc.

There is also a very broad room for development for deep learning algorithms. For example, how to make the algorithm run faster and reduce time consumption at the hardware and software levels; how to enhance the interpretability of the model to provide more help for future factor mining; how to make predictions more accurate, etc.

on deep learning quantitative finance is an interdisciplinary research topic in the field of international computer science and financial technology. The academic value and significance of the research proposed should be widely recognized and highly concerned by the international academic community.

**Annex 1 CNE5 model style factor description:**

| 因子类型 | 说明 |
|---|---|
| Beta 贝塔因子 | Beta 采用的是指数加权移动平均法来反应个股相对于市场整体的波动情况。 $$r_t - r_{ft} = \alpha + \beta R_t + e_t$$ |
| Momentum 动量因子 | 动量因子表示相对强弱的一个指标，通过超额对数收益的加权和计算得出。 $$Momentum = \sum_{t=L}^{T+L} w_t * [\ln(1+r_t) - \ln(1+r_{ft})]$$ 其中，T 为交易日，L 表示滞后阶数，Wt 通过指数加权移动平均法得出。 |
| Size 规模因子 | 公司股票总市值的对数。 |
| Earnings Yield 收益率因子 | 因子构成：0.68 * EPIBS + 0.11 * ETOP + 0.21 * CETOP 其中，EPIBS 为分析师预测的 E/P ratio；ETOP 为过去 12 个月的收益额与目前市值之比，其中过去 12 个月的收益额=上年财报收益额+（当前时点收益额-上年同时点收益额）；CETOP 为过去 12 个月的现金收益与当前现金价格之比。 |
| Residual Volatility 残差波动因子 | 残差波动因子由 DASTD（日标准差）、CMRA（累计收益边际）、HSIGMA（历史 Sigma）构成。 DASTD 是指数加权移动平均法计算过去交易日日超额收益率的波动率。 CMRA 是过去 12 个月内的最大累计超额收益与最小累计超额收益之差：$CMRA = Z_{max} - Z_{min}$。其中 $Z = \sum_{t=1}^{T} \ln(1+r_t) - \ln(1+r_{ft})$。 HSIGMA 表示残差收益的波动率，即不能 Beta 进行解释的额外收益的波动率：$\sigma = std(et)$，残差波动率因子需要与 Beta 因子进行回归消除共线性影响。 |
| Growth 成长因子 | 构成：0.47 * SGRO + 0.24 * EGRO + 0.18 * EGIBS + 0.11 * EGIBS_s 其中别是 SGRO 表示营业收入增长率、EGRO 表示利润增长率、EGIBS 表示长期预测利润增长率、EGIBS_s 表示短期预测利润增长率。 |
| Book-to-Price 账面价值与市值之比因子 | 表示去年财报普通股权的账面价值与当前普通股权的市值之比。 |
| Leverage 杠杆因子 | 构成：0.38 * MLEV + 0.35 * DTOA + 0.27 * BLEV 其中 MLEV 是市值杠杆、DTOA 是资产负债率、BLEV 是账面价值杠杆。 |
| Liquidity 流动性因子 | 构成：0.35 * STOM + 0.35 * STOQ + 0.30 * STOA 其中 STOM 是月换手率、STOQ 是季换手率、STOA 是年换手率。 |
| Non-linear size 非线性规模因子 | 这个因子为 Size（市值因子）的立方值，这里最终采用的结果是经过处理的，处理的过程为：首先，取 Size 的立方；然后将其与 Size 因子数值进行加权回归消除取共线性影响；最后对该因子进行标准化处理。 |